\documentclass[a4paper,11pt]{article}
\pdfoutput=1 
\usepackage{amsmath}
\usepackage{amssymb}
\usepackage{graphicx,graphics}
\usepackage{pdfpages}

\begin{document}

\noindent {\sc
14$^{\text{th}}$ Topical Seminar on Innovative Particle and Radiation Detectors (IPRD16) \\
3 - 6 October 2016\\
Siena, Italy
}

\vspace{1cm}

{\LARGE\flushleft\sffamily\bfseries\noindent DIAPHANE~: Muon tomography applied to volcanoes, civil engineering, archaelogy \par}
\flushbottom

\title{\boldmath DIAPHANE: Muon tomography applied to volcanoes, civil engineering, archaelogy}

\vspace{0.5cm}
\hrule height 1.5pt
\vspace{1.5cm}

\noindent {\bf Jacques Marteau$^{1,*}$, Jean de Bremond d'Ars$^2$, Dominique Gibert$^{2,3}$, Kevin Jourde$^{1}$, Jean-Christophe Ianigro$^{1}$, Bruno Carlus$^{1}$}\\

\noindent 1- {\it Institut de Physique Nucl\'eaire de Lyon (CNRS UMR 5822), Universit\'e de Lyon, Universit\'e Claude Bernard, Lyon, France.}\\

\noindent 2- {\it OSUR - G\'{e}osciences Rennes (CNRS UMR 6118), Universit\'{e} Rennes 1, Rennes, France.}\\

\noindent 3- {\it National Volcano Observatories Service, Institut de Physique du Globe de Paris (CNRS UMR 7154), Paris, France. }\\

\noindent (*) Corresponding author, {\it marteau@ipnl.in2p3.fr, www.diaphane-muons.com}\\

\noindent \textbf{Keywords} {Cosmic muons, Muography, Monitoring, Data processing methods, Instrumentation and methods for time-of-flight (TOF) spectroscopy}

\abstract{Muography techniques applied to geological structures greatly improved in the past ten years. Recent applications demonstrate the interest of the method not only to perform structural imaging but also to monitor the dynamics of inner movements like magma ascent inside volcanoes or density variations in hydrothermal systems. Muography time-resolution has been studied thanks to dedicated experiments, e.g. in a water tower tank. This paper presents the activities of the DIAPHANE collaboration between particle- and geo-physicists and the most recent results obtained in the field of volcanology, with a focus on the main target, the Soufri\`ere of Guadeloupe active volcano. Special emphasis is given on the monitoring of the dome's inner volumes opacity variations, that could be inferred to the hydrothermal system dynamics (vaporization of inner liquid water in coincidence with the appearance of new fumaroles at the summit). I also breifly present results obtained in the fields of civil engineering (study of urban underground tunnels) and archaelogy (greek tumulus scanning).  
}

\vfill

\pagebreak

\vspace{0.5cm}
\hrule height 1.5pt
\tableofcontents
\vspace{0.5cm}
\hrule height 1.5pt
\vspace{0.5cm}

\section{Introduction}  

Using the muon component of secondary cosmic rays to radiography geological bodies like volcano lava domes has greatly improved over the past ten years. Muon radiography, or muography, relies on the same principles than X-ray radiography, recovering the target density distribution, $\rho$, by measuring their screening effect on an incident particle beam, here the secondary cosmic muons. This approach is now detailed in many papers (see for example \cite{marteau2012muons} and references therein). Muography techniques are non-invasive, allow to operate relatively far from the target, and do not require repeated point-like measurements like e.g. gravimetry or electrical tomography \cite{marinita2016}. These features make them really adapted to the study of volcanic domes with $\sim$-kilometer diameters. This typical scale is set by the fact that the hard muon component is able to cross some kilometres of rock \cite{tanaka2001development, lesparre2010geophysical, lesparre2012density, jourde2013experimental, jourde_joined_inversion, jourde_nature_2016}. Applications to civil engineering (tunnels, dams) and environmental studies (near surface geophysics) are subject to active research, and monitoring of density changes in the near surface constitutes an important objective in hydrology and soil sciences.

In short, muography is sensitive to the target opacity (in $[\mathrm{g.cm}^{-2}]$ or in centimetres water equivalent $[\mathrm{cm.w.e.}]$), $\varrho$, which quantifies the amount of matter encountered by the muons along their path, $L$
\begin{equation}
\varrho = \int_L \rho(l) \times \mathrm{d}l.
\label{opacity1}
\end{equation}

The muons energy loss processes, typically $2.5 \; \mathrm{MeV}$ per $\mathrm{g.cm}^{-2}$, allow to predict a priori the quantity of muons absorbed in the target, given a minimal set of assumptions on the target constitution and outer boundaries and a reasonable modelling of the muon flux. Above several hundredths of $\mathrm{GeV}$, the typical energy cut-off imposed by kilometer-scale standard rock targets, the incident flux of muons may reasonably be considered stationary, azimuthally isotropic and to only depend on the zenith angle \cite{gaisser1990cosmic, lesparre2010geophysical}. The situation is quite different for targets with lower opacity whose studies require corrections for atmospheric effects \cite{jourde_monitoring_2016}, and to a lesser extend for geomagnetic effects \cite{munakata2000precursors}.

\section{The DIAPHANE project}  

The DIAPHANE collaboration gathers experts of HEP (IPN Lyon, France, CNRS/IN2P3, Universit\'e de Lyon) and geophysics (OSU Rennes and IPG Paris, France, CNRS/INSU, Universit\'es de Rennes and Paris). The collaboration started in 2008 to assess the feasibility of an active volcanic dome muography in the Lesser Antilles, the Soufri\`ere of Guadeloupe (ANR Domoscan). The muography conducted on the dome was the first in France. The muography study of the Lesser Antilles volcanoes has been extended by a second national funding in 2014 (ANR Diaphane). In parallel of this main project, which I detail in the following, different subprojects were undertaken~:
\begin{itemize}

\item other active volcanoes have been studied~: 
\begin{itemize}
\item the Etna south crater in 2012 (Italy, Sicily, \cite{carbone_experiment_2014} and 
\item the Mayon in the Philippines in 2014 (experiment still underway).
\end{itemize}

\item dedicated methodological measurements were conducted on various fields~:
\begin{itemize}
\item muography structural imaging of the Mont-Terri anticlinal from the Mont-Terri underground lab (Jura, Switzerland) since 2009 \cite{lesparre_mise_2011};
\item joined muon-gravimetry structural imaging of the Mont-Terri anticlinal since 2014 \cite{jourde_joined_inversion};  
\item online monitoring of opacity variations in a controlled environment (water tank) and measurement of the barometric corrections in 2015 \cite{jourde_monitoring_2016};  
\item structural imaging of a fault from the underground Tournemire laboratory (Aveyron, France) in collaboration with the IRSN \cite{lesparre2016cerdon};
\item Croix-Rousse civil underground tunnel scanned in Lyon (France) in 2015;
\item scanning of a greek tumulus in the Thessaloniki area \cite{gomez2016arche}.
\end{itemize}

\end{itemize}

\begin{figure}[htbp]  
\begin{center}
\includegraphics[width=0.95\linewidth,height=0.6\linewidth]{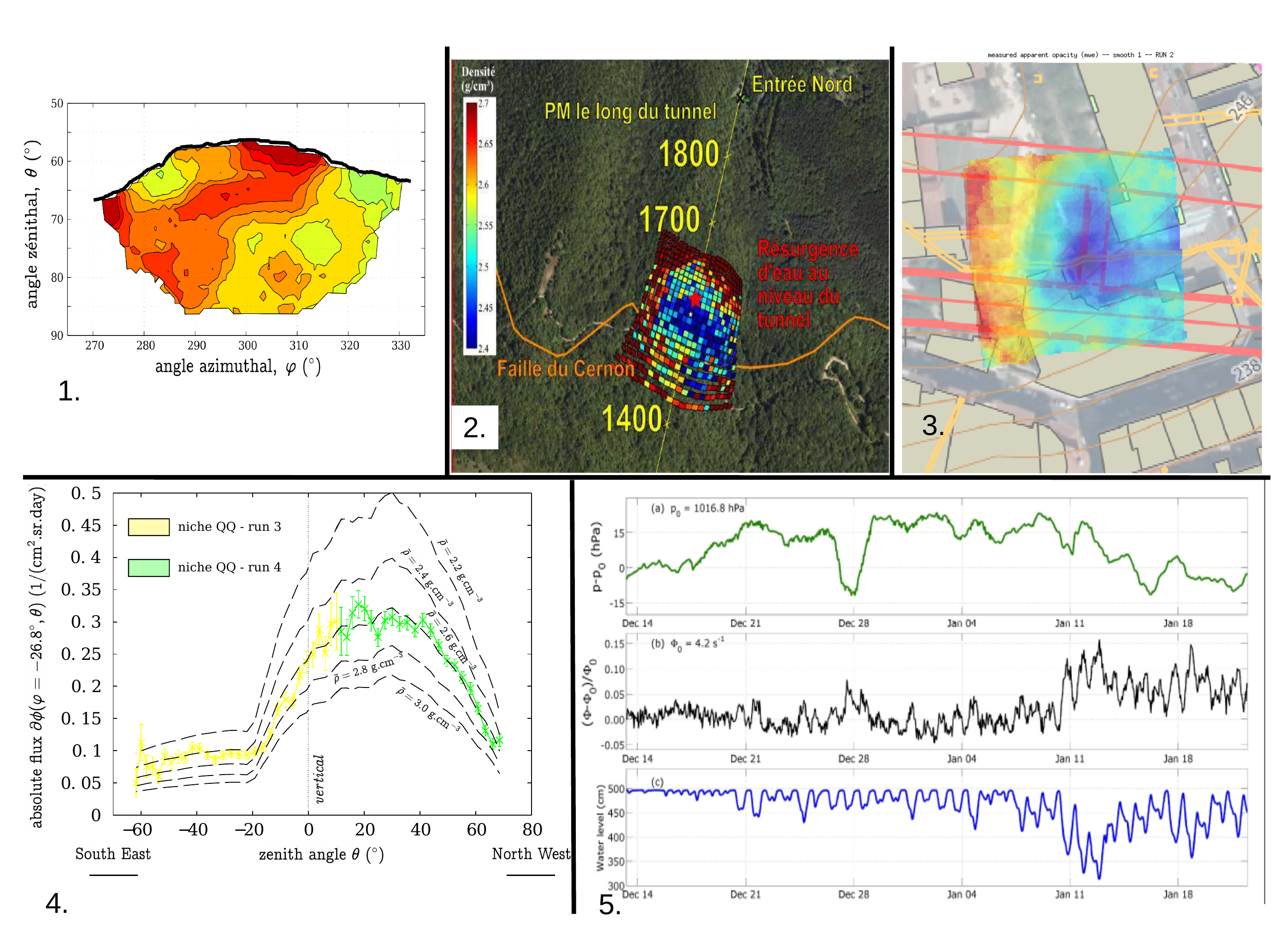}
\end{center}
\caption{\small
DIAPHANE structural imaging and monitoring results in various fields. \textit{1~:} Structural imaging of the La Soufri\`ere of Guadeloupe dome. \textit{2~:} Structural imaging of the Cerdon fault from the Tournemire underground laboratory. \textit{3~:} Structural imaging of the Croix-Rousse hill in Lyon, from the urban civil underground tunnel. \textit{4~:} Muon fluxes measured in the Mont-Terri underground laboratory (Switzerland) from 2 different acquisition sites. \textit{5~:} Online monitoring of the opacity variations in a water tank.}
\label{fig_all_results}
\end{figure}

The main results obtained are collected in the one-shot-view figure\ref{fig_all_results}. Beyond basic structural imaging the muography techniques are powerful 

\section{A field instrument}  

Muon detectors used in muography applications are multi-planes trackers with simple event-building features. The so-called ``direct'' problem, i.e. the flux measurement of the muons having crossed the target, consists in recording in time the direction of the muons incident on the detector. The main parameter is the acceptance function $\mathcal{T}_i \; [\mathrm{cm}^2 \cdot \mathrm{sr}]$ which relates the muon count, $N_i$, to the flux of cosmic muons, $\partial\phi \; [\mathrm{s}^{-1}\cdot \mathrm{cm}^{-2}\cdot \mathrm{sr}^{-1}]$ received in its $i^\mathrm{th}$ line of sight,
$$
N_i = T \times \int_{4\pi} \mathcal{P}_i (\varphi,\theta) \times \partial\phi (\varrho,\varphi,\theta) \times \mathrm{d} \Omega,
  = T \times \mathcal{T}_i \times \partial\phi_i,
$$
where $T$ is the duration of the acquisition period, $\mathcal{P}_i \; [\mathrm{cm}^2]$ is the detection surface function of the line of sight. Notice that $\partial\phi$, the differential flux of muons that reaches the instrument after crossing the target, depends both on the open sky differential flux $\partial\phi (\varrho = 0,\varphi,\theta)$ \cite{sagisaka1986atmospheric, ambrosio1997seasonal, adamson2010observation, poirier2011periodic, tang2006muon,heck1998corsika, wentz2003simulation} and on the muon absorption law inside matter. 

\begin{figure}[htbp]  
\begin{center}
\includegraphics[width=0.95\linewidth,height=0.6\linewidth]{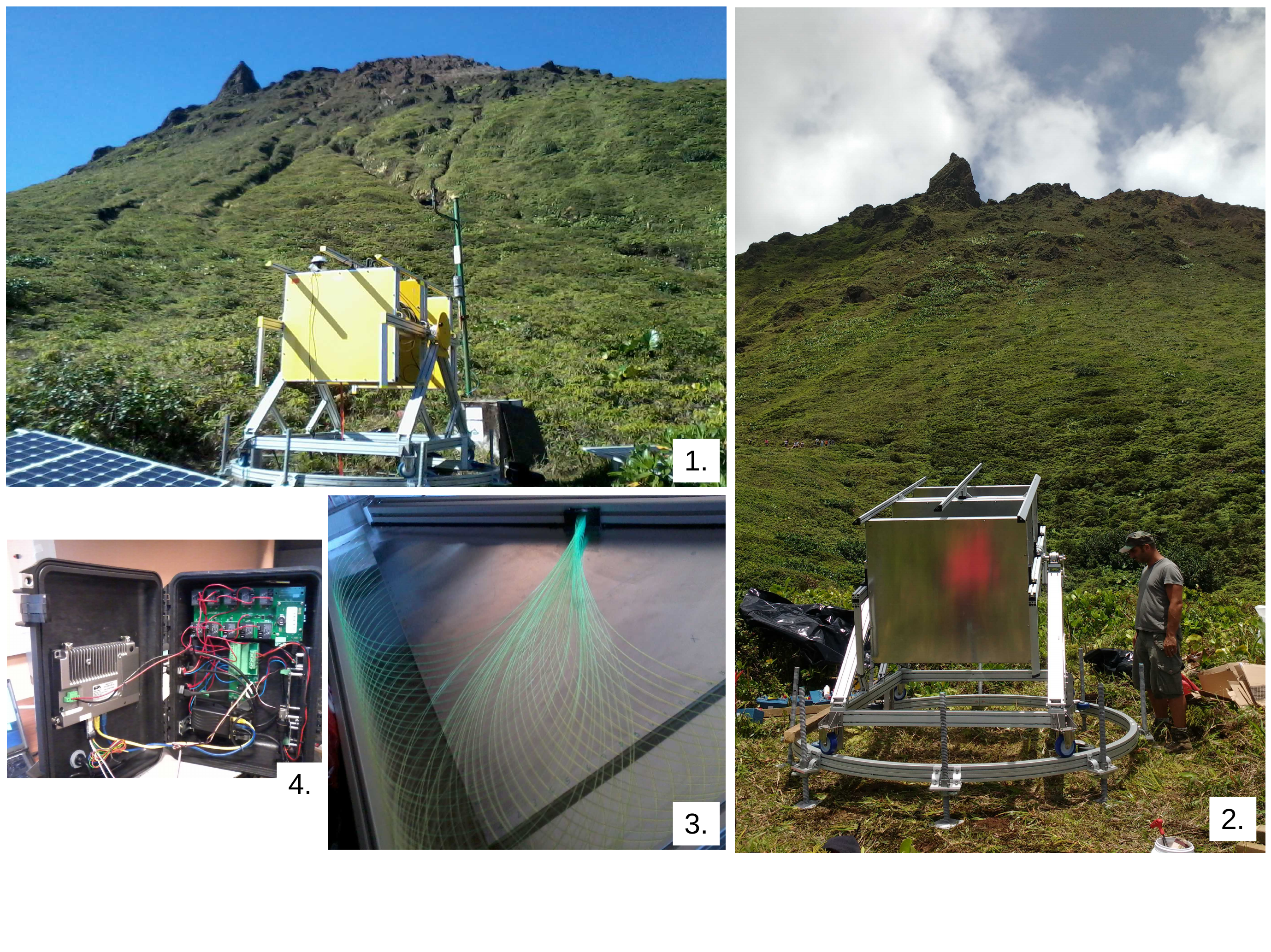}
\end{center}
\caption{\small
DIAPHANE muon detectors upgrades. \textit{1~:} The first generation 3 planes muon detector on the slope of La Soufri\`ere of Guadeloupe (PK site). \textit{2~:} The second generation 3 planes muon detector, with a transverse segmentation divided by a factor 2, on the slope of La Soufri\`ere of Guadeloupe (SAM site). \textit{3~:} Inner WLS fibres collected on a PMT cookie. \textit{4~:} Compact CTRL BOX with embedded hardened processing unit and electronics~: common clock signal, webrelay, Ethernet switch, POE to the wifi antenna.}
\label{all_telescopes_fig}
\end{figure}

The DIAPHANE detectors \cite{lesparre2012design, jourde2013experimental} are plastic scintillator-based, where the muons energy loss in the scintillator is converted into photons, shifted by WLS fibres, and brought down to multianodes PMT. The opto-electronics chain has been developed from HEP experiments on the concept of the autonomous, Ethernet-capable, low power, smart sensors \cite{marteau_opera_2010, marteau_implementation_2014}. These choices have been driven from the very beginning by the field constraints (volcanic domes without any infrastructures) imposing that the detectors should be not only sensitive but also robust, modular, transportable by all possible means, low consumption, autonomous etc. These choices were not led by the desire of spin-off R\&D's like in other projects and they allowed to perform real reasurements in real conditions practically at the startup of the project in 2008. Those detectors may be easily tuned on demand with the possible addition of extra detection planes, the change of the transverse segmentation, of the detection surface etc. Some pictures of the DIAPHANE detectors evolution are displayed in (Fig.\ref{all_telescopes_fig}).

Those detectors have a typical experimental acceptance function which maximum value is obtained for the line of sight perpendicular to the detector planes, corresponding to $(x,y) = (0,0)$. This acceptance function is determined experimentally to account for the detection matrices defects potentially induced by imperfect optical coupling(s). These defects cause distorsions in the acceptance function that factorize out when it is used for the muography data analysis. In practice, the determination of the acceptance is performed by measuring the open-sky muons flux at the zenith.
 
\section{Focus on the La Soufri\`ere of Guadeloupe dome~: structural and functional imaging}  

The volcanologic part of the project focuses on the Lesser Antilles, a subduction volcanic arc with a dozen of active volcanoes, such as the Montagne Pel\'ee in Martinique, the Soufri\`ere of Guadeloupe, and Soufriere Hills in Montserrat, which all presented eruptive activity during the 20$^\mathrm{th}$ century. The Soufri\`ere of Guadeloupe is an active volcano which last important manifestation was a phreatic eruption in 1976-1977. The present dome is very young ($\sim$ 500 years)  \cite{Boudon2008, Komorowski2008} and sits on a 15$^o$ N-S inclined plane  leading to an unstable structure and is very heterogeneous, with massive lava volumes embedded in more or less hydrothermalized materials \cite{lefriant2006} which evolve quickly with time because of the constant and massive erosion induced by the heavy tropical rains \cite{gibert2010muon}. All these features and the dome's dimensions, typically $\sim$ 500m in height and radius, makes the La Soufri\`ere of Guadeloupe volcano an extremely valuable target for muography investigation, in particular to try to constrain the occurence of the most likely hazards for la Soufri\`ere today, a new phreatic eruption or a flank collapse.

\begin{figure}[htbp]  
\begin{center}
\includegraphics[width=0.95\linewidth,height=0.6\linewidth]{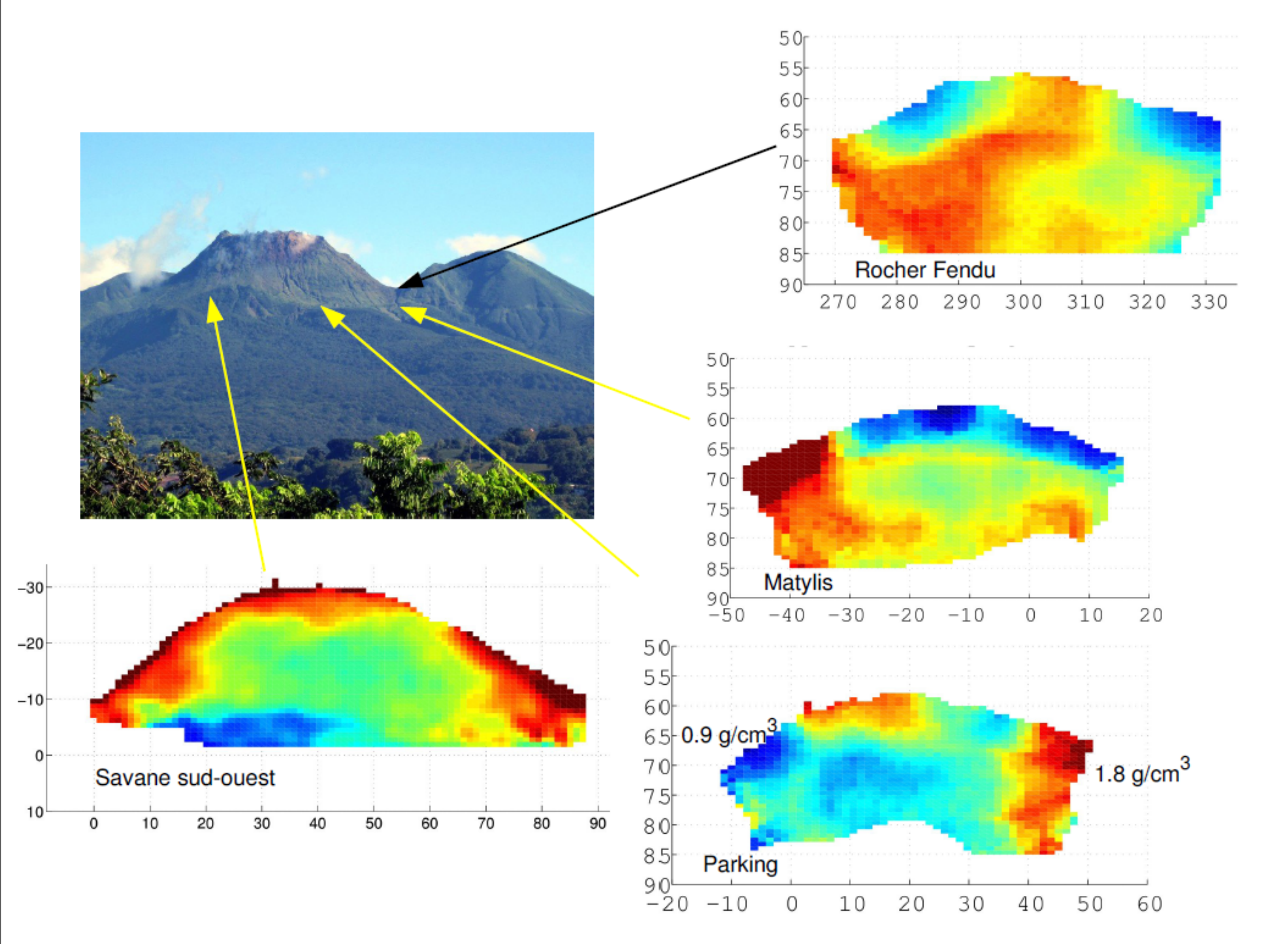}
\end{center}
\caption{\small
Structural imaging of the La Soufri\`ere of Guadeloupe dome from 4 different acquisition sites around the dome. The blue areas are the less dense zones of teh volcano. The red areas have the highest density. Average density extracted from all those images ranges from 1.6 to 1.8 $\mathrm{g.cm}^{-3}$.}
\label{fig_all_radios_souf}
\end{figure}  

The structural imaging of the La Soufri\`ere of Guadeloupe dome has been performed by the DIAPHANE collaborators in the 2010-2016 period from 5 different acquisition sites. 4 of them are around the dome (corresponding to the images displayed in Fig.\ref{fig_all_radios_souf}) and the last one corresponds to a dedicated smaller detector placed in a fault below the South crater of the volcano. In 2010-2014 the same detector was moved from place to place, while since 2015 5 detectors are continuously taking data. The images are corrected from the upward-going particles flux which could mimic through-going particles close to the horizon \cite{jourde2013experimental, marteau_implementation_2014}. The excellent quality of those images reveals the heterogeneity of the dome, with low density regions in particular, either close to the surface, corresponding to the most active craters and fumaroles at the summit (see image from Rocher Fendu and Matylis sites), or within the inner volume of the dome corresponding to large empty volumes (see image from Parking site). Comparisons with standard geophysical imaging methods, like electrical tomography, show a good agreement of the muography data \cite{marinita2016}, which have the advantadge of being much simpler to invert.

\begin{figure}[htbp]  
\begin{center}
\includegraphics[width=0.8\linewidth]{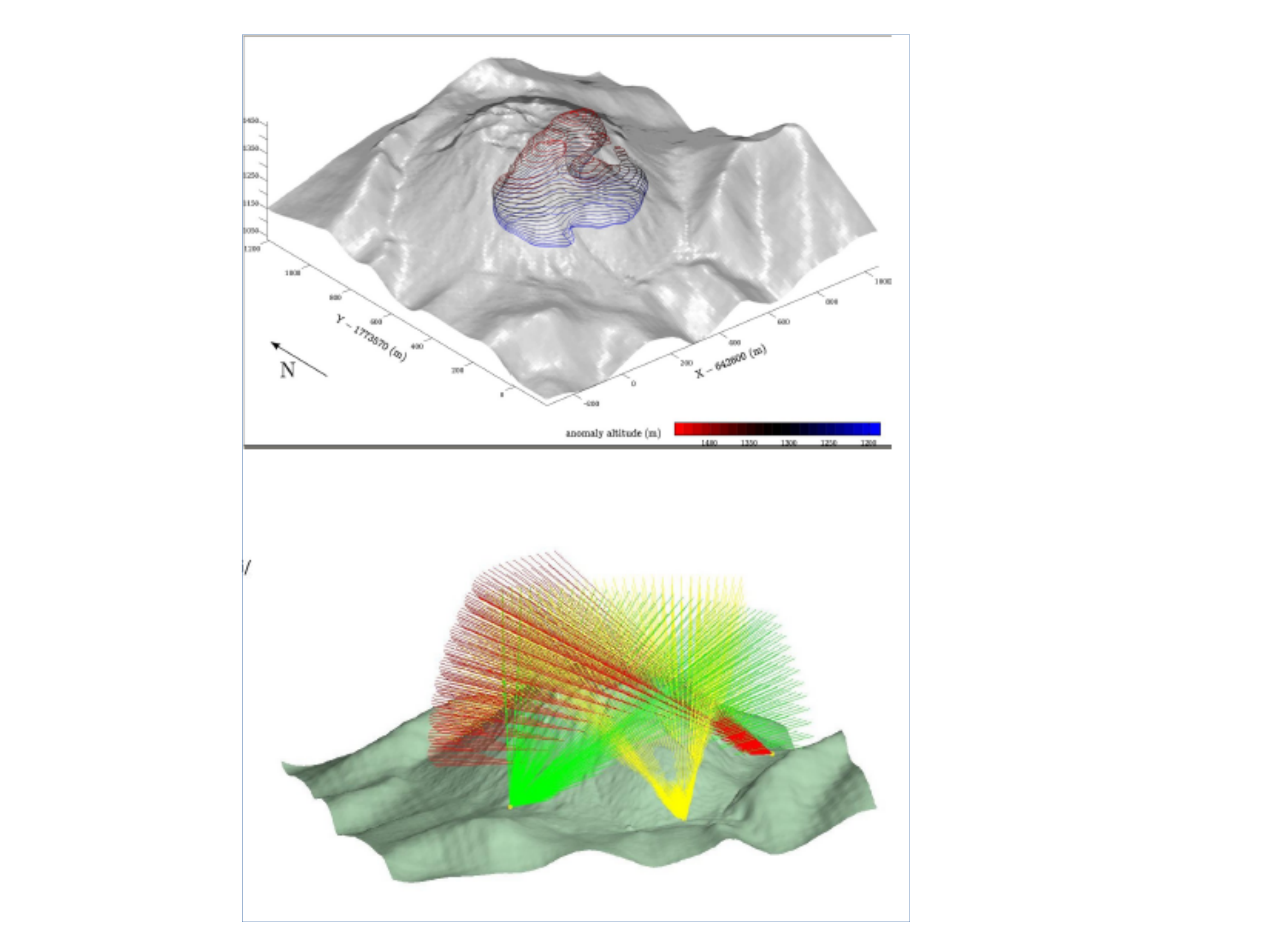}
\end{center}
\caption{\small
3D structural imaging of the La Soufri\`ere of Guadeloupe dome obtained after a joined inversion including gravimetry data (top). Muons trajectories from 4 different acquisition sites (bottom).}
\label{fig_tomo_all_views_3D}
\end{figure}

Those 2D images acquired from different sites may be combined to reconstruct the dome structure in 3D (from radiography to tomography). In a joined analysis melting muography and gravimetry data we obtained a 3D map of the dome, as displayed in Fig.\ref{fig_tomo_all_views_3D}.   
 
\begin{figure}[htbp]  
\begin{center}
\includegraphics[width=0.95\linewidth,height=0.6\linewidth]{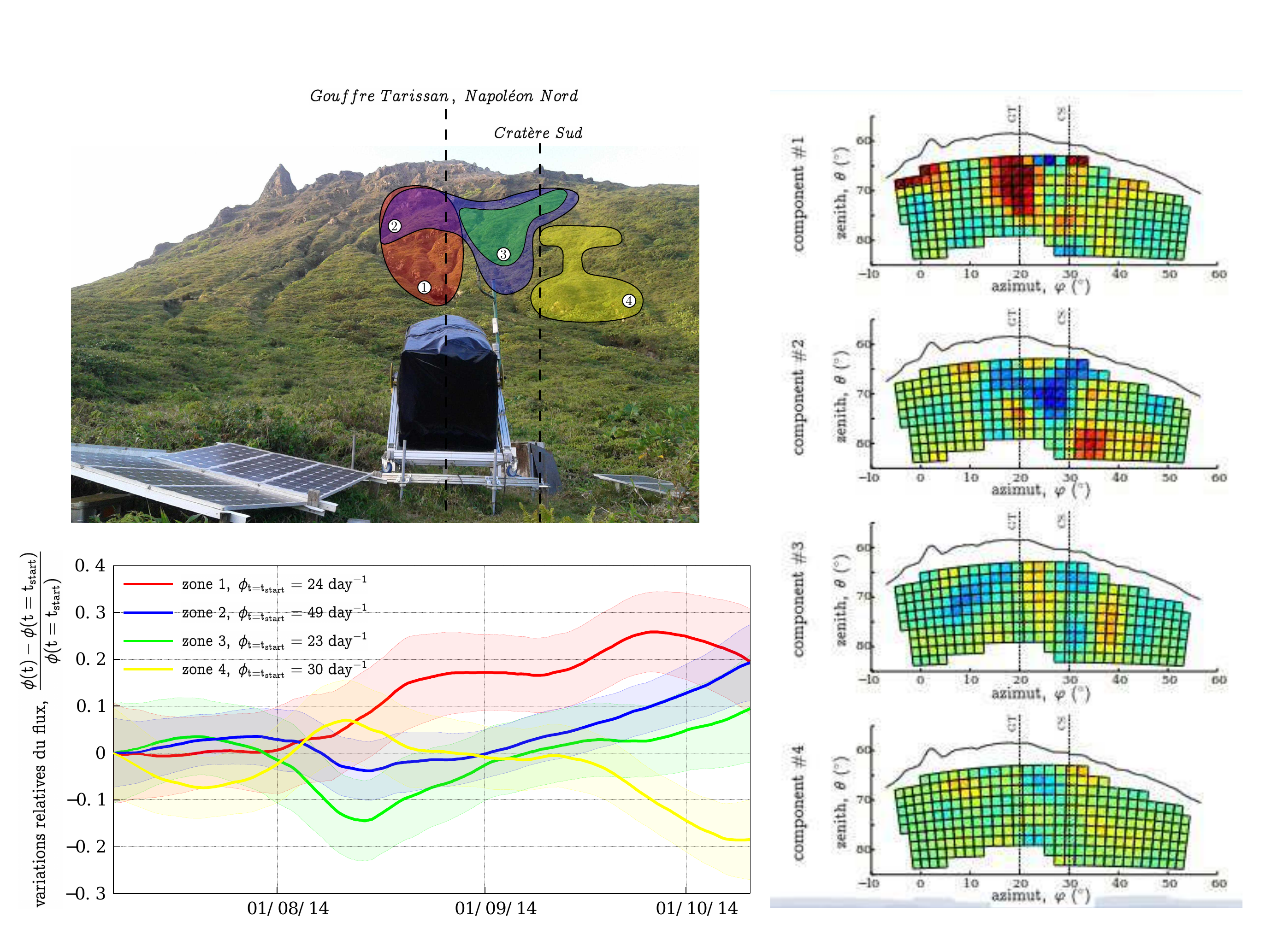}
\end{center}
\caption{\small
Functional imaging of the La Soufri\`ere of Guadeloupe dome. Left top~: picture of the muon detector facing the dome. Projected on the surface of the dome, the 4 different areas which coherent time variations (left, bottom). Right~: principal vector components analysis defining of the 4 areas.}
\label{fig_all_monitoring}
\end{figure}

Beyond the structural imaging capabilities of the muography, the DIAPHANE collaboration also proved, by studying long data time series, that the method is sensitive to the dynamics of the fluids movements inside the target \cite{jourde_nature_2016}. Indeed muons flux exhibits different time behaviours for different lines of sight. In Fig.\ref{fig_all_monitoring} are displayed the main results of the Summer 2014 data analysis corresponding to the appearance of new fumaroles at the summit of the volcano. From the time distributions it appears clearly that the muons crossing the various areas 1 to 4 have different variations in time, which can be inferred neither to open-sky muon flux variations (which would impact all lines of sight) nor to instrumental effects, for the same reason. A principal vector components analysis allows to gather lines of sight with coherent time behaviour. The measurements are compatible with the volcanologic observations of a significant increase of the dome fumarolic activity since 2014 \cite{allard_steam_2014, ovsg_bilan_2015, ovsg_bilan_2016}. Indeed a new active region appeared to the North-East of the Tarissan pit during the 2014 Summer \cite{ovsg_bilan_2015}, the North-Napoleon fumarole, and two old pits, the Gouffre Breislack and the Gouffre 56, have seen their activity rising \cite{ovsg_bilan_2016}. The volumes of these domains vary from $1 \times 10^6 \; \mathrm{m}^3$ to $7 \times 10^6 \; \mathrm{m}^3$. The estimated masses contained in those volumes are compatible with a total mass budget which remains approximately constant~: two domains show a mass loss ($\Delta m \in [-0.8 ; -0.4] \times 10^9 ~ \mathrm{kg}$) and the third one a mass gain ($\Delta m \in [1.5 ; 2.5] \times 10^9 ~ \mathrm{kg}$). The negative mass changes may be inferred to the formation of steam in shallow hydrothermal reservoir previously partly filled with liquid water. 

This last result allows to open a new field for the muography techniques, beyond the standard structural imaging. The ability of the method to provide online monitoring information reinforces its interest in geosciences applications, in complement with standard techniques, also sensitive to the density distributions, which may be jointly inverted with muography data to further constrain the geophysical models. 

\paragraph{Acknowledgments}   
Most of the results obtained are part of the DIAPHANE project ANR-14-CE 04-0001. We thank the IPNL technical team, L.Germani, J.-L.Montorio, F.Mounier.
This work is done in memory of Fabrice Dufour. This is IPGP contribution ****.



\end{document}